\documentclass[graybox]{svmult}

\usepackage{amsmath}
\usepackage{amssymb}
\usepackage{mathptmx}       
\usepackage{helvet}         
\usepackage{courier}        
\usepackage{type1cm}        
\usepackage{makeidx}         
\usepackage{graphicx}        
\usepackage{multicol}        
\usepackage[bottom]{footmisc}

\makeindex             

\begin{document}

\title*{The Efficiency of Coherent Radiation from Relativistic Shocks}
\titlerunning{The Efficiency of Coherent Radiation from Relativistic Shocks}
\author{Takanobu Amano, Masanori Iwamoto, Yosuke Matsumoto, and Masahiro Hoshino}
\institute{
Takanobu Amano \at
Department of Earth and Planetary Science, University of Tokyo, 7-3-1, Hongo, Bunkyo-ku, Tokyo, 113-0033, Japan
\email{amano@eps.s.u-tokyo.ac.jp}
\and
Masanori Iwamoto \at
Department of Earth and Planetary Science, University of Tokyo, 7-3-1, Hongo, Bunkyo-ku, Tokyo, 113-0033, Japan
\email{iwamoto@eps.s.u-tokyo.ac.jp}
\and
Yosuke Matsumoto \at
Department of Physics, Chiba University, 1-33, Yayoi, Inage-ku, Chiba, Chiba, 263-8522, Japan
\email{ymatumot@chiba-u.jp}
\and
Masahiro Hoshino \at
Department of Earth and Planetary Science, University of Tokyo, 7-3-1, Hongo, Bunkyo-ku, Tokyo, 113-0033, Japan
\email{hoshino@eps.s.u-tokyo.ac.jp}
}
%
%
\maketitle

\newcommand{\abstracttext}{
We discuss a mechanism for intense electromagnetic wave emission at an astrophysical relativistic shock \index{relativistic shock} in a magnetized collisionless plasma \index{collisionless plasma}. At the magnetized shock, the particle reflection by a compressed magnetic field of the shock produces a ring-like distribution in momentum, which gives rise to plasma instabilities. Intense and coherent high-frequency electromagnetic waves will be emitted if the synchrotron maser instability (SMI) \index{synchrotron maser instability} is excited, whereas non-propagating magnetic fluctuations will be generated when the Weibel instability \index{Weibel instability} (WI) is the dominant mode. The problem is of great astrophysical interest because if intense radiation is emitted, the interaction with the upstream medium induces a large-amplitude electrostatic field (or Wakefield), which may play a role for the acceleration of ultra-high-energy cosmic rays \index{cosmic rays}. We review our recent effort to measure the efficiency of the electromagnetic wave emission using fully self-consistent, two-dimensional (2D) particle-in-cell (PIC) \index{particle-in-cell} simulations for pair plasmas. We found that the emission efficiency in 2D was systematically lower than one dimensional (1D) PIC simulation results. However, the power remains finite even when the WI is active to generate large-amplitude magnetic fluctuations. Astrophysical implications of the present results are briefly discussed.
}

\abstract*{\abstracttext}
\abstract{\abstracttext}

\section{Introduction} \label{sec:intro}
Powerful low-frequency electromagnetic radiations are often observed associated with collective space and astrophysical plasma dynamics \cite{Melrose2017CoherentPlasmas}. An extremely efficient electromagnetic wave emission may result from groups of charged particles moving in phase with each other, just like a single particle that carries a much larger number of charges. In such coherent radiation \index{coherent radiation}, the emitted power is proportional to $N^2$ (where $N$ is the number of charged particles moving in phase) rather than $N$ in incoherent radiation. Therefore, the coherent radiation is much stronger than the incoherent radiation,  provided that such coherent charged particle bunches are generated and maintained for a sufficiently long time in natural collisionless plasma environments in space. Examples of such coherent radiation sources inferred from observations include planetary magnetospheres (emission from auroral regions), solar corona (solar radio bursts), pulsars (pulsed emission in radio).

Although there has been no observational identification, a relativistic shock propagating in a magnetized collisionless plasma may also be a source of strong coherent radiation. A magnetized collisionless shock involves the reflection of particles coming into the shock by a sharp increase in the magnetic field at the shock front. The reflected particles gyrate around the upstream magnetic field and are on average accelerated by traveling along the direction of the upstream motional electric field. Within the shock transition layer where both the reflected and incoming populations coexist, the momentum distribution of the particles should appear as a ring-like distribution in the plane perpendicular to the local magnetic field. Such a ring-like distribution may become unstable against several different types of plasma instabilities. In the case of a highly relativistic shock, the synchrotron maser instability (SMI) \cite{Hoshino1991PreferentialPlasmas} may be excited from a relativistic ring distribution of electrons (or positrons). The instability results from the resonance between an electromagnetic wave of extraordinary mode (X-mode) and the $n$-th harmonics of the cyclotron motion of the relativistic particles. One may write the resonance condition as $\omega = n \omega_{ce}$ where $\omega$ is the electromagnetic wave frequency and $\omega_{ce}$ is the relativistic cyclotron frequency of an electron. Since the frequency of an X-mode electromagnetic wave is always higher than the cyclotron frequency, higher harmonics resonance $n > 1$ is needed in general for the instability. The emission of X-mode electromagnetic waves via the SMI at magnetized relativistic shocks was demonstrated by earlier studies using fully self-consistent Particle-In-Cell (PIC) simulations in one dimension (1D) \cite{Langdon1988StructurePlasmas,Hoshino1992RelativisticPositrons,Gallant1992RelativisticPlasmas}. Being the first principles approaches for numerical modeling of collisionless plasmas, the PIC simulations require extensive computational resources and two-dimensional (2D) and three-dimensional (3D) simulation were not possible in the old days. Meanwhile, the same unstable distribution function may excite the Weibel instability (WI) \cite{Weibel1959} if the ambient magnetic field is sufficiently small. Since the mode is unstable for the wave vector perpendicular to the shock normal, multidimensional simulations are essential to reproduce this instability.

With ever-increasing computational capabilities, recent PIC simulations of collisionless shocks have been performed routinely in 2D and sometimes even in 3D \cite{Sironi2011PARTICLESHOCKS,Sironi2013THESHOCKS,Matsumoto2017}. Some earlier multidimensional PIC simulation studies of magnetized relativistic shocks reported that the coherent precursor waves were not observed, and suggested that the coherent radiation found in 1D may be an artifact of the reduced dimensionality. However, one has to be careful in interpreting the results because the numerical resolutions used in the earlier studies might not be sufficient. We expect that the high-frequency and short-wavelength electromagnetic radiation is sensitive to the choice of the mesh size and may quickly be damped at low resolutions. Furthermore, a PIC simulation involving a relativistic bulk flow becomes unstable against a numerical Cherenkov instability. A typical strategy is to use a digital filtering technique to eliminate unphysical short-wavelength fluctuations, which would also suppress physical electromagnetic waves even if they present. These numerical issues have been an obstacle that makes it difficult to estimate the emission efficiency accurately.

Concerning astrophysical applications, such intense radiation from a relativistic shock drew attention as it may play a key role for the acceleration of cosmic rays (CRs) possibly to ultra-high energies \cite{Chen2002,Lyubarsky2006ApJ,Hoshino2008}. The origin of CRs has been a long-standing puzzle in astrophysics, in particular at its highest energy part or Ultra-High-Energy Cosmic Rays (UHECRs). The first order Fermi acceleration, which remains as the leading mechanism for the CR acceleration, has a difficulty at relativistic magnetized shocks because of the inefficiency of particle diffusion across the magnetic field line. Therefore, an alternative model for the CR acceleration at a relativistic magnetized shock is of great interest to the astrophysics community.

Consider an intense and coherent electromagnetic wave packet propagating in an electron-proton plasma. It pushes electrons via the ponderomotive force \index{ponderomotive force} (or a wave pressure) in the direction of the wave propagation. The protons, on the other hand, are left behind because of the much larger inertia. A charge separation will then develop, which produces a large-amplitude electrostatic field called a Wakefield. It propagates with the group velocity of the wave packet, which is very close to the speed of light. A particle trapped in an electrostatic potential may then be accelerated linearly by the propagating wave packet \cite{Tajima1979LaserAccelerator}. This mechanism, called the Wakefield acceleration \index{Wakefield acceleration}, has been known as an efficient particle acceleration mechanism in the context of the laser-plasma interaction but may provide a means to accelerate CRs if indeed a relativistic shock emits intense electromagnetic radiation.

The relatively new scheme for the CR acceleration has a number of advantages over the standard first-order Fermi acceleration. Nevertheless, the model is not yet sophisticated enough especially because we do not know the intensity of radiation emitted from a relativistic shock.

In this paper, we present our recent 2D simulation results that accurately quantified the coherent wave emission efficiency at relativistic magnetized shocks in electron-positron (pair) plasmas. In particular, we investigated the dependence on the upstream magnetization parameter defined as the ratio between the Poynting flux to the particle kinetic energy flux. The results suggest that the emission efficiency in general is lower than in 1D, but nevertheless always remains finite for a parameter range we have investigated. Astrophysical implications of the obtained results are briefly discussed. More detailed discussion may be found in our recent publications \cite{Iwamoto2017,Iwamoto2018}.

\section{Simulations} \label{sec:simulation}
\subsection{Method and Setup}
The PIC scheme is the standard numerical method for studying collisionless plasma dynamics in a fully self-consistent manner. It deals with the motions of super-particles in continuous phase space. The electromagnetic field is defined on a mesh and typically advanced in time using a finite difference scheme. The charge and current densities at the mesh points are calculated by taking velocity moments of the particle distribution function and used for updating the electromagnetic field by solving Maxwell's equations. The Lorentz forces acting onto the particles may be obtained by interpolating the electromagnetic fields defined at neighboring cells.

We used a 2D PIC simulation code \cite{Matsumoto2015,Ikeya2015StabilitySimulations} that employs a magic CFL (Courant-Freidrichs-Levy) number to minimize the effect of a numerical Cherenkov instability. The performance of the code concerning the suppression of the numerical instability was investigated in detail by \cite{Ikeya2015StabilitySimulations}. In this study, we only present 2D simulation results with a fixed numerical resolution, but the resolution dependence was checked by performing a separate numerical convergence study in a 1D simulation box \cite{Iwamoto2017}.

We used an injection method to generate a shock wave in the simulation box. Namely, particles with both positive and negative charges are injected from one side of the box with a relativistic bulk velocity. The injected particles are reflected at the other side of the box. The reflected and incoming particles interact with each other, which results in the particle heating and deceleration of the incoming bulk flow. Because of the symmetry, the plasma bulk flow becomes zero as a result of this heating. When an injection speed is supersonic, a shock wave forms in between the freshly injected (the upstream) plasma and the heated plasma (the downstream). The shock then propagates toward the incoming plasma to satisfy the Rankine-Hugoniot relationships. The simulation frame thus corresponds to the downstream rest frame.

We take the $x$ direction to be parallel to the shock normal and place the injection and reflection boundaries at the upper and lower bounds in $x$, respectively. We performed the simulations in the $x{\rm -}y$ plane and assumed that everything is initially homogeneous in the $y$ direction. Accordingly, the boundary condition is periodic in the $y$ direction.

In this study, we only consider a magnetized perpendicular shock, which is defined as a shock with the upstream ambient magnetic field perpendicular to the shock normal (or parallel to the shock surface).  We may thus arbitrarily choose the ambient magnetic field in the $y{\rm -}z$ plane. Because of the 2D simulation box, the shock dynamics, in general,  is dependent on the choice of the upstream magnetic field direction even if all the other parameters are fixed. More specifically, if the magnetic field is in the $z$ direction or the out-of-plane configuration, the particle gyromotion is contained in the simulation plane, while the motion parallel to the magnetic field line is not appropriately taken into account. On the other hand, if it is taken to be in the $y$ direction or the in-plane configuration, the situation becomes opposite. It is difficult to predict the differences (if any) in between the two cases without actually performing the simulations. Therefore, we performed simulations both with the in-plane and out-of-plane configurations. In this way, we investigated possible artifacts of the 2D assumption and tried to infer the fully 3D situation as much as possible.

Our primary motivation in this study is to obtain a quantitative estimate of the efficiency of electromagnetic radiation and also understand the physic behind. Specifically, we investigated the emission efficiency at a relativistic shock in an electron-positron plasma to reduce the computational cost. We note that this choice makes it impossible to investigate the efficiency of the Wakefield acceleration itself as it requires a finite inertia difference between positive and negative charges. Nevertheless, high-frequency electromagnetic radiation is produced only by leptons. We thus think that the emission efficiency measured in this study will be a reasonable estimate even for an electron-proton plasma. This conjecture should be checked by performing simulations for an electron-proton plasma in the future.

We used a fixed injection Lorentz factor of $\gamma_1 = 40$ in all the simulations presented in this paper. The emission efficiency was then measured as a function of the magnetization parameter:
\[
\sigma_e = \frac{B_1^2}{4 \pi \gamma_1 N_1 m_e c^2}
= \left( \frac{\omega_{ce}}{\omega_{pe}} \right)^2
\]
for both the in-palne and out-of-plane configurations independently. The dependence on $\sigma_e$ is important for astrophysical applications because it is common to use the magnetization parameter to characterize the properties of relativistic jets. Note that $\omega_{pe} = \sqrt{4\pi N_1 e^2 / \gamma_1 m_e}$, $\omega_{ce} = e B_1 / \gamma_1 m_e c$ are the electron plasma and cyclotron frequencies; $B_1$ and $N_1$ are the ambient magnetic field strength and number density in the far upstream region. The notations for the other quantities are standard. Note that we use the CGS units throughout in this paper.

A cell size of $\Delta x / c/\omega_{pe} = 1/40$, the number of particle per cell in the upstream of $N_1 \Delta x^2 = 64$ were used with the number of cells of $(N_x, N_y) = (20,000, 1680)$ for the $x$ and $y$ directions, respectively.

\subsection{Shock Structures} \label{subsec:structures}

\begin{figure}[t]
\begin{center}
\includegraphics[scale=0.40]{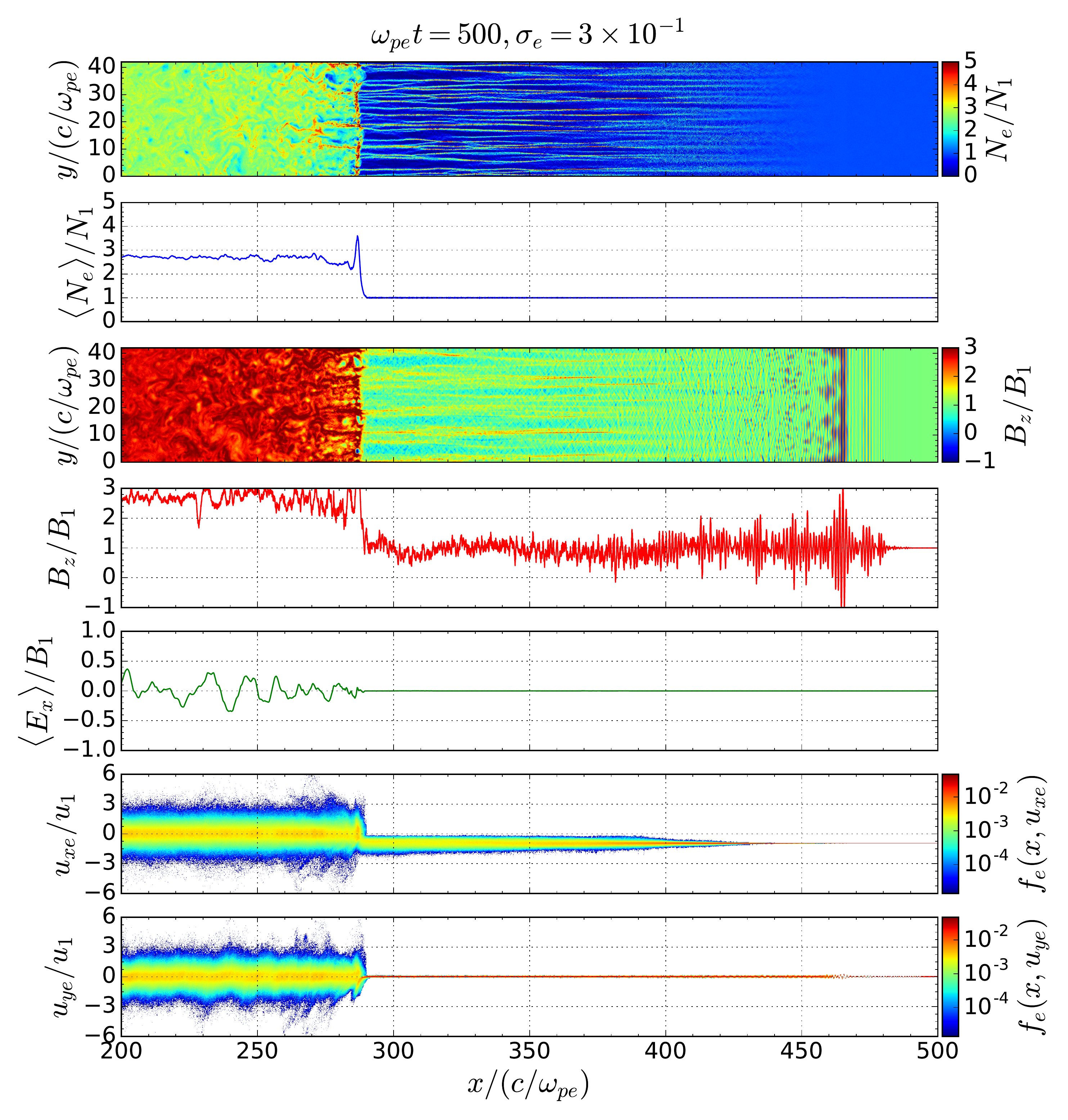}
\caption{Shock structure for $\sigma_e = 3 \times 10^{-1}$ with the out-of-plane configuration at $\omega_{pe} t = 500$ (after \cite{Iwamoto2017}). From top to bottom, the electron density, $y$-averaged electron density, magnetic field $B_z$, slice of magnetic field along $x$, $y$-averaged electric field $E_x$, phase space plots in the $u_{x}{\rm -}x$ and $u_{y}{\rm -}x$ space for electrons, respectively.}
\label{fig:01}
\end{center}
\end{figure}

We first discuss the simulation results obtained with the out-of-plane magnetic field configuration. The overall shock structure for $\sigma_e = 3 \times 10^{-1}$ is shown in Fig.~\ref{fig:01}. The upstream plasma on the right-hand side traveling toward the left is compressed at a well-developed shock at $x/c/\omega_{pe} \simeq 290$, which may be identified by a sharp density increase. One observes large-amplitude ($\delta B/B_0 \sim 1$), and short-wavelength magnetic fluctuations in the upstream region of the shock. We confirmed that these are the precursor electromagnetic waves, which were generated at the shock and propagating toward the upstream with a group velocity greater than the shock speed.

Another finding in this simulation is the clear filamentary structures in density in the precursor region. We note that the filaments are generated well ahead of the shock transition region, and thus not a result of the WI (which is not active at this moderate magnetization). Therefore, the formation of the filaments should be attributed to a nonlinear interaction between the intense electromagnetic waves and the upstream plasma. Although we have not yet fully understood the formation mechanism, a nonlinear wave-wave coupling (or a parametric instability \index{parametric instability}) of the large-amplitude precursors is likely to be the cause. The formation of the density filaments thus indicates that the emitted electromagnetic waves are strong and coherent.

Note that one may identify the heating of the incoming plasma already in the deep precursor $x/c/\omega_{pe} \sim 300{\rm -}350$. The emission efficiency at the shock should then be modified because it is the pre-heated plasma that enters into the shock and excites the instability. We confirmed that the shock structure and the precursor wave amplitude become almost stationary at this stage. In the following, we measured the wave power well after such a quasi-steady state has been established. Therefore, the nonlinear feedback effect of the wave emission to the shock itself has already been taken into account.

\begin{figure}[t]
\begin{center}
\includegraphics[scale=0.40]{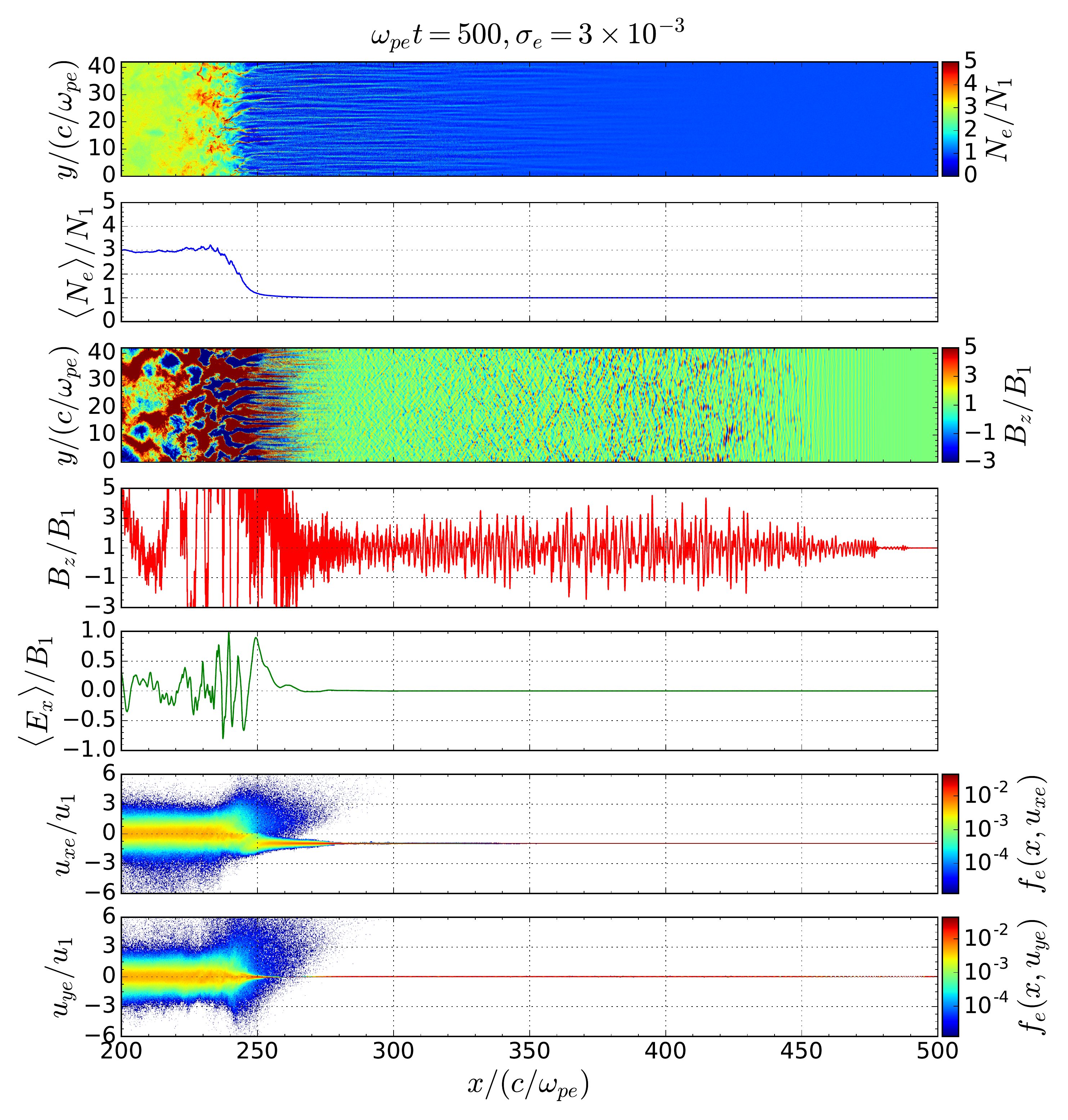}
\caption{Shock structure for $\sigma_e = 3 \times 10^{-3}$ with the out-of-plane configuration at $\omega_{pe} t = 500$ (after \cite{Iwamoto2017}). The format is the same as Fig.~\ref{fig:01}.}
\label{fig:02}
\end{center}
\end{figure}

In Fig.~\ref{fig:02}, the simulation results obtained with $\sigma_e = 3 \times 10^{-3}$, again with the out-of-plane configuration, is shown with the same format. Again, we observed the large-amplitude precursor waves and density filaments. In addition, the lower ambient magnetic field strength (or low $\sigma_e$) activates the WI in the shock transition region $x/c/\omega_{pe} \sim 250$. Note that the amplitude of Weibel-generated magnetic field is much larger than the upstream magnetic field, but the color for the $B_z$-panel saturates in the shock transition region because we chose a color scale that emphasizes the precursor wave emission. It is known that the WI generates filamentary structures both in density and magnetic field. However, the density filaments are generated already in the far upstream region where the reflected particles (with positive $u_x$) were not seen. This indicates that the filamentary structure in the precursor is generated by the nonlinear interaction between the precursor waves and the incoming plasma rather than the WI, as in the case of the moderate magnetization $\sigma_e = 3 \times 10^{-1}$. In other words, the simulation result demonstrates the emission of intense electromagnetic waves from the shock that is largely dominated by the Weibel-generated turbulent magnetic field.

\begin{figure}[t]
\begin{center}
\includegraphics[scale=0.40]{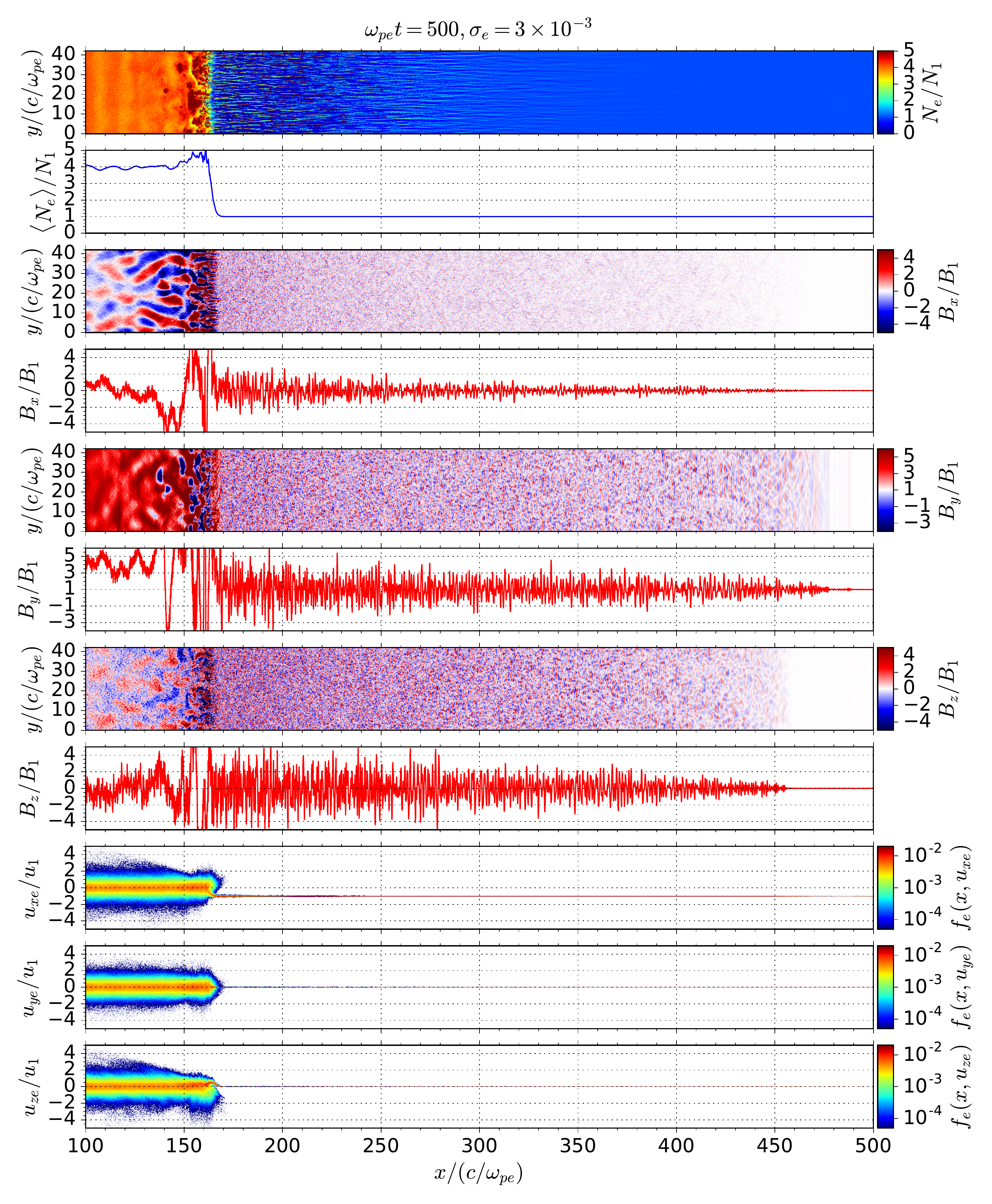}
\caption{Shock structure for $\sigma_e = 3 \times 10^{-3}$ with the in-plane configuration at $\omega_{pe} t = 500$ (after \cite{Iwamoto2018}). From top to bottom, the electron density, $y$-averaged electron density, $B_x$, 1D slice of $B_x$ along $x$, $B_y$, 1D slice of $B_y$ along $x$, $B_z$, 1D slice of $B_z$ along $x$, phase space plots in the $u_x{\rm -}x$, $u_y{\rm -}x$, and $u_z{\rm -}x$ space for electrons, respectively.}
\label{fig:03}
\end{center}
\end{figure}

We now present the results for the in-plane magnetic field configuration. Shown in Fig.~\ref{fig:03} is the result obtained with the same $\sigma_e = 3 \times 10^{-3}$, but with the in-plane configuration. We again observed the clear precursor wave emission and the filamentary structures in density. Since the same magnetization $\sigma_e$ was used, it is natural that the WI is prominent in the shock transition region.

We note that the precursor electromagnetic waves were found both in $y$ and $z$ components of the magnetic field. This is surprising because the linear theory of the SMI predicts that the wave magnetic field should be polarized in the direction of the ambient magnetic field (waves on the X-mode dispersion brunch). Indeed, the waves were polarized only in the $z$ direction for the out-of-plane configuration. Similarly, one would naturally expect the polarization is in the $y$ direction in the in-plane configuration. The waves polarized in the $z$ direction are therefore unexpected. As we discuss later, we think that the waves of unexpected polarization were emitted associated with the Weibel-generated large-amplitude magnetic fluctuations in the shock transition region.

\subsection{Emission Efficiency} \label{subsec:efficiency}

\begin{figure}[t]
\begin{center}
\includegraphics[scale=0.6]{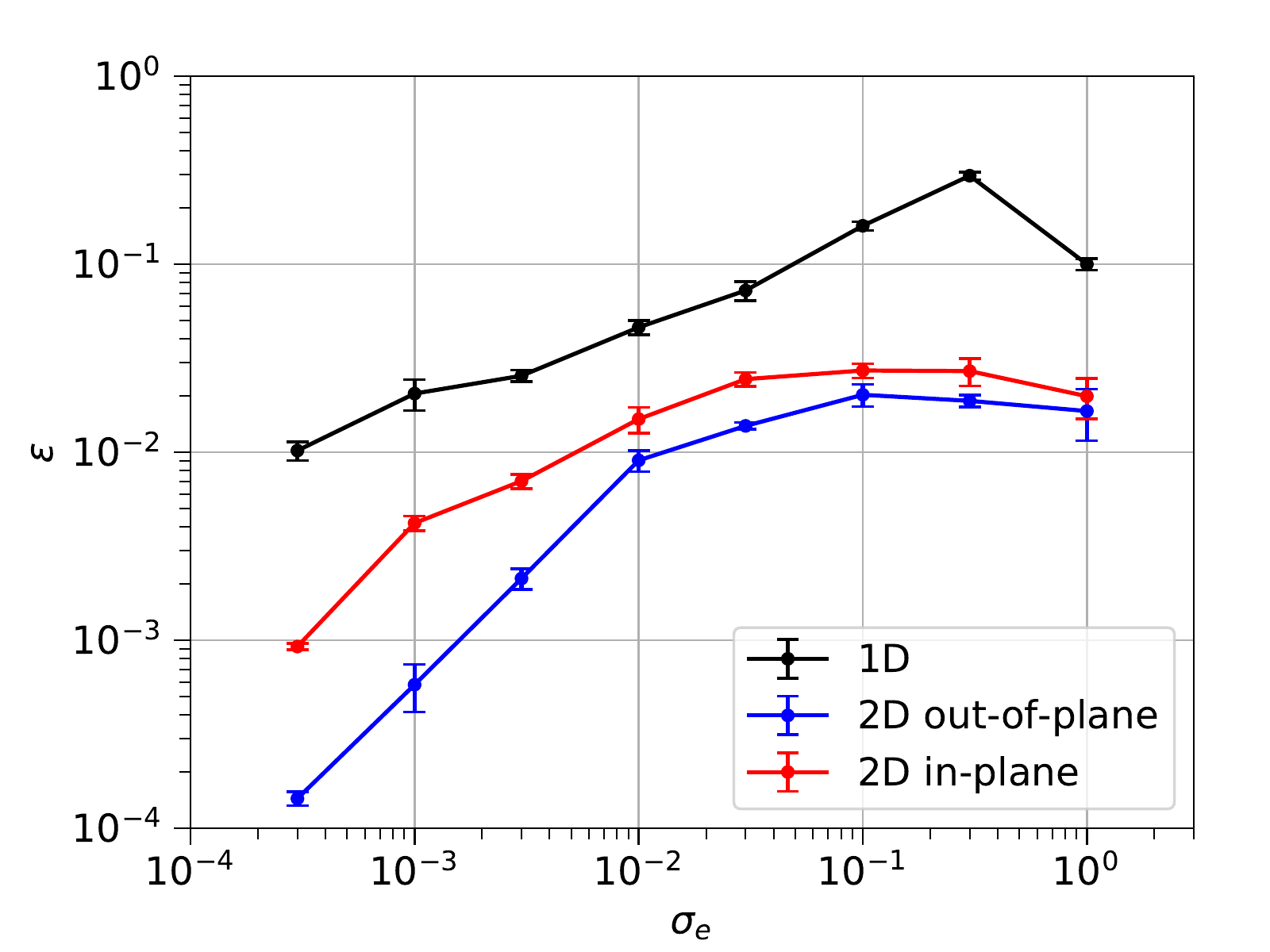}
\caption{Emission efficiency $\epsilon = \delta B^2/8 \pi \gamma_1 N_1 m_e c^2$ as a function $\sigma_e$. The black, blue, and red lines denote 1D, 2D with the out-of-plane, and 2D with the in-plane configurations, respectively.}
\label{fig:04}
\end{center}
\end{figure}

Figure \ref{fig:04} displays the compilation of our simulation results. The precursor wave power normalized to the upstream flow kinetic energy $\epsilon = \delta B^2/8 \pi \gamma_1 N_1 m_e c^2$ is shown as a function of magnetization $\sigma_e$. Since the upstream flow kinetic energy provides the free energy for the SMI and the resulting electromagnetic wave emission, one may interpret $\epsilon$ as the energy conversion efficiency. Note that we included both $y$ and $z$ components of the magnetic field fluctuations in the estimate of the wave power.

At moderate magnetization $\sigma_e \sim 10^{-1}$, the energy conversion efficiency is approximately a few \% both for the in-plane and out-of-plane configurations. The efficiency was lower than the corresponding 1D results by about one order of magnitude. This reduction may be due to the feedback effect of the emission. The upstream plasma is pre-heated already far ahead of the shock, which induces the filaments in density. The pre-heating will reduce the growth rate, especially at high wavenumbers. Also, the filaments eventually interact with the shock and generate the inhomogeneity in the transverse direction at the shock. The inhomogeneity will reduce the coherence of the ring-like distribution and also the growth rate. Despite the reduction in the emission efficiency, the emitted power in 2D has yet remained strong enough to drive relativistic transverse motions of electrons \cite{Iwamoto2017,Iwamoto2018}.

At small magnetization $\sigma_e \lesssim 10^{-2}$, the efficiency for the out-of-plane configuration starts to decline rapidly with decreasing the magnetization $\sigma_e$. This reduction may be attributed to the presence of the WI, which will strongly deteriorate the coherence of the ring-like distribution because of large-amplitude fluctuations $\delta B/B_1 \gg 1$. Nevertheless, it is important to point out that the WI does not entirely kill the precursor wave emission. The simulation results suggest that both instabilities coexist at least to some extent.

The reduction of the efficiency was less pronounced in the in-plane than in the out-of-plane configurations. This difference is partly due to the waves of unexpected polarization (or $\delta B_z$). Indeed, the contribution of this component was larger in the small magnetization regime where the WI dominates the shock transition region. This result suggests that the WI does not necessarily prohibit the precursor wave emission, but may, in principle, enhance the efficiency. Another possible reason is that the unperturbed gyromotion of the particles, which occurs in the $x{\rm -}z$ plane, is less affected by the turbulence in the this geometry. Nevertheless, as decreasing the magnetization, the large-amplitude turbulent magnetic field $\delta B_z/B_1 \gg 1$ in the shock transition region will make this effect relatively unimportant.

The discrepancy at low $\sigma_e$ found in between the two magnetic field configurations clearly indicates that fully 3D simulations are needed for a more quantitative and accurate estimate. Nevertheless, the emission efficiency is maintained at a relatively high level in both cases. Therefore, we believe that the emission of intense precursor waves is indeed an intrinsic nature of a magnetized relativistic shock.

\section{Summary and Discussion} \label{sec:discussion}
We have performed 2D PIC simulations of magnetized relativistic shocks in pair plasmas. The results have demonstrated that the emission of intense electromagnetic radiation reported previously with 1D PIC simulations is not an artifact of the low dimensionality. At least, shocks with magnetization $\sigma_e \gtrsim 10^{-4}$ emit strong electromagnetic precursor waves.

The critical parameter for the Wakefield acceleration model is the so-called strength parameter defined as $a = e E / m_e c \omega$, where $E$ is the wave electric field: an efficient particle acceleration has been observed in 2D PIC simulations with a simplified setup when the injected electromagnetic wave pulse satisfies $a > 1$. By assuming the scaling law $a \propto \gamma_1 \sqrt{\epsilon}$, Iwamoto et al. (2018) estimated a condition for $\gamma_1$ and $\sigma_e$ that needs to be satisfied for the emitted wave to be strong enough $a > 1$. The conclusion was that the Wakefield acceleration at a highly relativistic shock favorably with a moderate magnetization remains as a possible mechanism for the acceleration of UHECRs. In this scenario, highly relativistic jets from gamma-ray bursts are the most plausible sites of the production of UHECRs.

Since both the SMI and WI grow from the same free energy source, we anticipated that the precursor wave emission would cease if the growth rate of the WI dominates over the SMI. The simulation results have clearly shown that this is not the case. The reason for this may be understood by the small-scale nature of the turbulence generated by the WI. The Lorentz force of the random turbulent magnetic field $\delta B$ as seen from a relativistic particle is proportional to $\delta B \lambda/c$ where $\lambda$ is the coherence length of the turbulence. The factor $\lambda/c$ indicates the transit time of a particle going through the coherence length. On the other hand, the average Lorenz force due to the ambient magnetic field $B_0$ is proportional to $B_0/\omega_{ce}$. It is possible to show that the average Lorenz force is always greater than the random one for the Weibel-generated turbulence with the coherence length on the order of the electron skin depth $\lambda \sim c/\omega_{pe}$. Therefore, on average, a particle in the turbulence performs a complete gyromotion. We note that the growth rate of the SMI is on the order of the cyclotron frequency. Therefore, the coherent charge-particle bunches can develop during this gyromotion and eventually strong electromagnetic waves are emitted.

However, the presence of the WI is obviously not completely negligible. The turbulence gives random kicks in the coherent gyromotion of the particles, which gives rise to an effective temperature to the ring-like distribution in momentum. A finite temperature, in general, reduces the growth rate of the SMI in particular at short wavelengths or high harmonic numbers $n$. On the other hand, for the wave-particle resonance to occur with the brunch of high-frequency electromagnetic waves (so that the wave can escape upstream), the wave frequency should be larger than the plasma frequency $\omega = n \omega_{ce} \gtrsim \omega_{pe}$ (or the cut-off frequency in a strict sense). This leads to the condition $n \gtrsim \sigma_e^{-1/2}$ required for the harmonic numbers. As the magnetization $\sigma_e$ decreases, the heating by the WI will become more and more prominent, and the growth rates at high $n$ numbers will decrease. At the same time, the required harmonic numbers $n$ will become larger and larger. We thus expect that there will be a critical threshold in $\sigma_e$, below which the precursor wave emission will diminish. A much larger computational resource is needed to prove this conjecture.

It is important to mention that once a charged-particle bunch is generated, the magnetic field amplified by the WI may help to enhance the emission simply because the radiated power is proportional to the magnetic field energy density in the synchrotron theory. Our interpretation for the higher emission efficiency of the unexpected component ($\delta B_z$) in the in-plane configuration is that the polarization is locally of X-mode type with the perturbed magnetic field oscillates in the direction of the local magnetic field vector. Because of the large-amplitude turbulence in the shock transition region, the magnetic field may be locally in the $z$ direction. If a pre-existing charged bunch travels through a region of large $B_z$, a large-amplitude wave with the unexpected polarization may be emitted. The actual power of coherent radiation in the Weibel-dominated regime should be determined by the competition between the efficiency of charged-bunch generation and the magnetic field amplification. The answer to the question can only be given by conducting direct 3D simulations.

In this study, we employed an electron-positron plasma to estimate the emission efficiency. As we mentioned earlier, this makes sense because the emission will be generated only by electrons even at a relativistic electron-proton plasma shock. However, this implicitly assumes that the energy transfer between the two constituents does not occur. If the electrons are able to absorb energy from the protons, the efficiency should be rescaled by the effective kinetic energy available for the electrons. This would make the system more complex and therefore interesting. Simulation results for an electron-proton shock will be reported elsewhere in the future.

\begin{acknowledgement}
This work was supported in part by JSPS KAKENHI Grant Numbers 17H02966, 17H06140, 17H02877. This work used the computational resources of Cray XC30 and computers at Center for Computational Astrophysics, National Astronomical Observatory of Japan, the K computer provided by the RIKEN Advanced Institute for Computational Science, and the HPCI system provided by Information Technology Center, Nagoya University through the HPCI System Research Project (Project ID: hp150263, hp170158, hp180071).
\end{acknowledgement}

\bibliographystyle{spphys}
\bibliography{mendeley_v2}

\end{document}